\documentclass[12pt]{iopart}
\usepackage{graphicx}
\usepackage{color}
\usepackage[utf8]{inputenc}

\begin{document}
\title{Magnetic phase diagram of the Hubbard model
  with next-nearest-neighbour hopping}
\author{R.~Peters$^1$ and T. ~Pruschke$^{1,2}$}
\address{$^1$Department of Physics, University of G\"ottingen, Friedrich-Hund-Platz 1,
37077 G\"ottingen, Germany}
\address{$^2$The Racah Institute of Physics, Hebrew University of Jerusalem, Jerusalem 91904, Israel}
\ead{peters@theorie.physik.uni-goettingen.de}
\begin{abstract}
We calculate the magnetic phase diagram of the Hubbard model 
for a Bethe lattice with
nearest neighbour (NN) hopping $t_1$ and next nearest neighbour (NNN) hopping 
$t_2$ in the limit of infinite coordination.
We use the amplitude $t_2/t_1$ of the NNN hopping to tune
the density of states (DOS) of the non-interacting system from a situation with
particle-hole symmetry to an asymmetric one with van-Hove singularities
at the lower ($t_2/t_1>0$) respectively upper ($t_2/t_1<0$) band edge for large
enough $\left|t_2/t_1\right|$.
For this strongly asymmetric situation we find rather extended parameter
regions with ferromagnetic states and regions with antiferromagnetic states.
\end{abstract}

\submitto{\NJP}
\section{Introduction}
The phenomenon of spontaneous magnetism is one of the oldest topics in physics.
That lodestone can attract iron is known for over 2500 years. In contrast,
a rigorous understanding of the microscopic processes which lead to
magnetism still is a matter of present day research \cite{siegmann}.
In order to microscopically describe the phenomenon ``magnetism''
quantum mechanics, in particular the spin of the electrons,
and the inclusion of interactions respectively many-body correlations are
mandatory.

A further typical property of materials which show magnetic behaviour is
that they possess partially filled d- or f-shells. In this case, orbital
degrees of freedom usually quite dramatically influence the existence and
nature of magnetically ordered states. 
A rather notorious example are the manganites, which show a rather complex
phase diagram due to an interplay of orbital and spin degrees of freedom
\cite{coey1999,salamon2001}.

A much simpler situation occurs when, for example, in a crystal with low
symmetry due to lattice distortions, only one of the d- or f-states
effectively plays a role at the Fermi energy. In this case one can think
of an effective one-band model as appropriate description. A well-known example
for such a situation are the cuprate superconductors \cite{imada1998}.
Here, too, magnetic order can occur. However, while for materials with
orbital degrees of freedom the existence of both antiferromagnetism and
ferromagnetism can easily be accounted for \cite{imada1998}, the one-band situation
prefers the formation of antiferromagnetic order \cite{fuldebook}. While
ferromagnetic states are known to exist under certain extreme conditions
\cite{nagaoka1966}, their possible occurrence and stability regions in physically
relevant model parameter regimes is still an intensively investigated research
topic.

In this paper we therefore want to focus on the one-orbital situation.
A suitable model for describing strong correlation physics in such a single
band is the Hubbard model \cite{hubbard1963,kanamori1963,gutzwiller1963}
\begin{displaymath}
H=\sum_{ij,\sigma}t_{ij}c_{i\sigma}^\dagger
c_{j\sigma}-\mu\sum_{i\sigma}n_{i\sigma}+U\sum_i
n_{i\uparrow}n_{i\downarrow}\quad .
\end{displaymath}
The operator $c_{i\sigma}^\dagger$ creates an electron with spin
$\sigma$ at site $i$, $t_{ij}$ describes the ``hopping'' amplitude
from site $i$ to $j$ and $\mu$ is the chemical potential, which can be
used for tuning the occupation of the system. The two particle
interaction is purely local and
only entering  via a product of two density operators
$n_{i\uparrow}=c_{i\uparrow}^\dagger c_{i\uparrow}$  with amplitude
$U$. 

In recent years progress in
understanding the physics of this model in dimensions larger than one
was mostly gained from calculations using the dynamical mean field theory (DMFT)
\cite{georges1996,pruschke1995} or cluster variants of it
\cite{maier2005}. The DMFT relates the lattice model 
to an impurity model in an effective medium representing the lattice, which 
must be solved self-consistently. It can be shown that this mapping is
exact in the limit of infinite spatial dimensions or infinite coordination
of the lattice \cite{georges1996,metzner1989}.
Note that the remaining (effective) impurity problem represents a quantum-impurity,
which by itself is complicated to solve. From the methods available
we here use the numerical
renormalisation group (NRG) \cite{wilson1975,bulla2008}, because it
is by far the most efficient and accurate technique for single-band problems.
For the calculation of  spectral functions we employ the complete Fock space
variant \cite{peters2006,weichselbaum2007} of the NRG.

For real three dimensional materials the DMFT is, of course, only an
approximation. Nevertheless, the Hubbard model within DMFT describes a
lot of strong correlation physics, which can be seen in real
materials, at least qualitatively correct. In this sense it
is therefore justified to study for example magnetic properties of the
Hubbard model within this approximation. As the DMFT can be seen as a
thermodynamically consistent mean-field theory
\cite{georges1996,janis1992},
one can expect that the phase diagram obtained at least gives an account
for potential phases, albeit not necessarily the correct phase boundaries.

The aim of the present paper is to give an account of the possible
antiferromagnetic and ferromagnetic phases of the doped 
single-band Hubbard model.
For a particle-hole symmetric density of states (DOS) the model has
an antiferromagnetically ordered ground state at half filling for every
finite value of $U$, which phase separates upon doping 
\cite{dongen1994,dongen1995,zitzler2002}.
Ferromagnetism can also be found in the single band Hubbard model, but only
for very high interaction parameter 
and close to half
filling \cite{nagaoka1966,obermeier1997,zitzler2002,park2008},
or for a pronounced asymmetric DOS 
also for moderate values of $U$ \cite{wahle1998,ulmke1998}.

Deviations from particle-hole symmetry in the single-band model leading to
such asymmetries in the DOS are achieved by inclusion of longer-range 
single-particle hopping processes. It is important to stress that
in DMFT the actual lattice structure only enters via the DOS.
As we are interested in a qualitative
investigation of the possible magnetic phases, it is permissible to work
with a computationally convenient DOS,
which is the one obtained from an infinitely-coordinated
Bethe lattice \cite{georges1996} with nearest neighbour (NN) and next-nearest neighbour (NNN)
hopping
$t_1$ respectively $t_2$. For $t_2=0$ one obtains the well-known semicircular
DOS \cite{georges1996}, which for values $t_2>0$ becomes asymmetric and can even
develop a singularity at one of the band edges \cite{kollar2005,eckstein2005}. 
From this point of view, the Bethe lattice in the limit of infinite coordination
has all typical features of the DOS of a real lattice -- compact support, van-Hove singularities --
and one can hope that results obtained with it give a reasonable qualitative
account of true three-dimensional systems. 

The paper is organised as follows. In the next section we
introduce the DOS of the $t_1$-$t_2$ Bethe lattice with infinite coordination,
which will be used throughout the paper. Section three
focuses on the antiferromagnetic phase, which is realised
near half filling. In section four we present the results for the
ferromagnetic calculations. Quite surprisingly, for strong enough 
$t_2$ we observe regions, where both antiferromagnetic and ferromagnetic states are stable. A summary and discussion will conclude the paper.
\section{Density of States\label{sec:DOS}}
\begin{figure}[htp]
\begin{center}
\includegraphics[width=0.8\textwidth,clip]{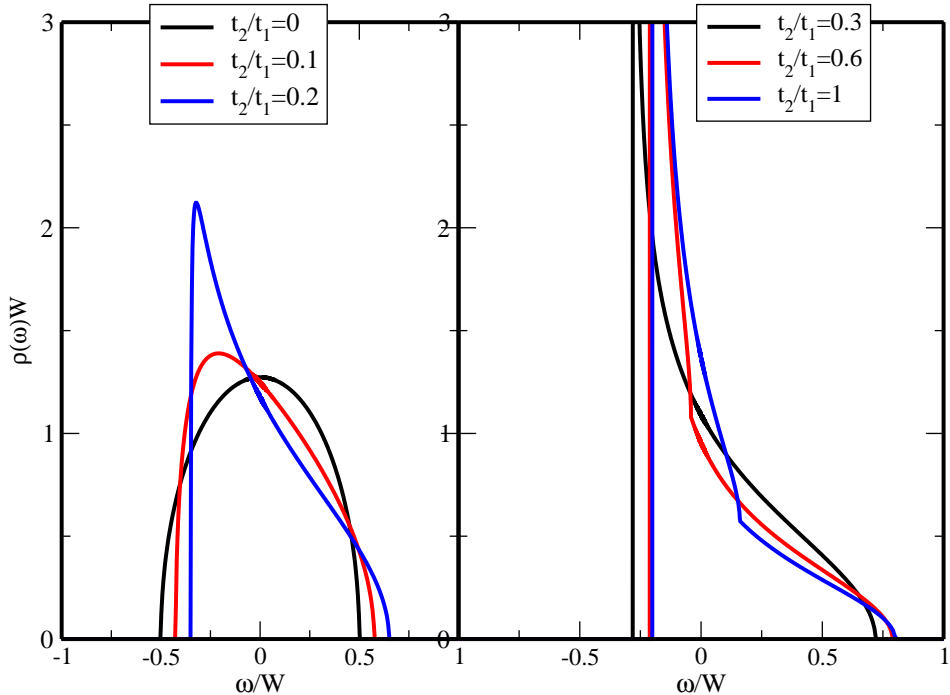}
\end{center}
\caption{Density of states for the Bethe lattice with NN and NNN
  hopping and different ratios $t_2/t_1$. The left side shows
  $t_2/t_1<0.25$, where no singularity at the lower band edge
  appears. The right side shows $t_2/t_1>0.25$ with a singularity at
  the lower band edge. The axis were scaled with the proper
  bandwidths. For $t_2/t_1<0$ the corresponding figures are
obtained by simply replacing $\omega\to-\omega$.}\label{DOS}
\end{figure}

Early studies of the Bethe lattice with longer-ranged hopping usually focused
on the simplified variant proposed by Georges et al.\ \cite{georges1996,chitra1999,zitzler2004}.
While in this approximation one introduces frustration to magnetic correlations, the resulting DOS
retains particle-hole symmetry, which of course is somewhat artificial.
The proper form of the DOS was deduced by Kollar et al.\ \cite{kollar2005,
eckstein2005}.
Figure \ref{DOS} shows the result
for different ratios of $t_2/t_1$. 
The non-interacting Green's function $G_{t_1,t_2}(\zeta)$ and by this the DOS
$\rho(\omega)=-1/\pi\Im G_{t_1,t_2}(\omega+i\eta)$ for the Bethe lattice with
nearest neighbour hopping $t_1$ and next nearest neighbour hopping
$t_2$ in the limit of infinite coordination is given by the formula
\begin{displaymath}
G_{t_1,t_2}(\zeta)=\frac{1}{2t_2b(\zeta)}\left[G_{t_1}\left(a+b(\zeta)\right)-G_{t_1}\left(a-b(\zeta)\right)\right]\;\;,
\end{displaymath}
with $a=\frac{-t_1}{2t_2}$,
$b(\zeta)=\sqrt{\frac{t_1^2}{4t_2^2}+\frac{\zeta}{t_2}+1}$ and
$G_{t_1}(z)=\frac{1}{2}\left(z-\sqrt{z^2-4}\right)$.
Analysing this
formula shows that there appears a singularity in the DOS for
$t_2>\frac{1}{4}t_1$. The singularity
is due to the factor $1/b$ and thus is a square root singularity. 
For $t_2<\frac{1}{4}t_1$ the band edges lie at $\omega_{1,2}=3t_2\pm 2t_1$. For
$t_2>\frac{1}{4}t_1$ the lower band edge is $\omega_1=-\frac{t_1^2}{4t_2}-t_2$
and the upper band edge $\omega_2=3t_2+2t_1$. Thus the bandwidth is
\begin{displaymath}
W=\left\{\begin{array}{lr}4t_1\qquad &t_2/t_1<1/4\\
2t_1+4t_2+t_1^2/(4t_2)\qquad &t_2/t_1>1/4
\end{array}\right.
\end{displaymath}
It should be emphasised that by tuning the NNN hopping $t_2$, the DOS
change from a particle-hole symmetric semi-ellipse to a strongly
asymmetric DOS with singularity for $t_2/t_1>\frac{1}{4}$.
This is a rather important feature expected to occur also in real materials.
On the other hand, previous investigations of frustration effects within DMFT
used the so-called
two sub-lattice fully frustrated model 
\cite{georges1996,duffy1997,hofstetter1998,chitra1999,zitzler2004}, which
misses this particular asymmetry and the van-Hove singularity.
\section{Magnetic phases close to half filling}
\subsection{$t_2=0$}
\begin{figure}[htp]
\begin{center}
\includegraphics[width=1\textwidth,clip]{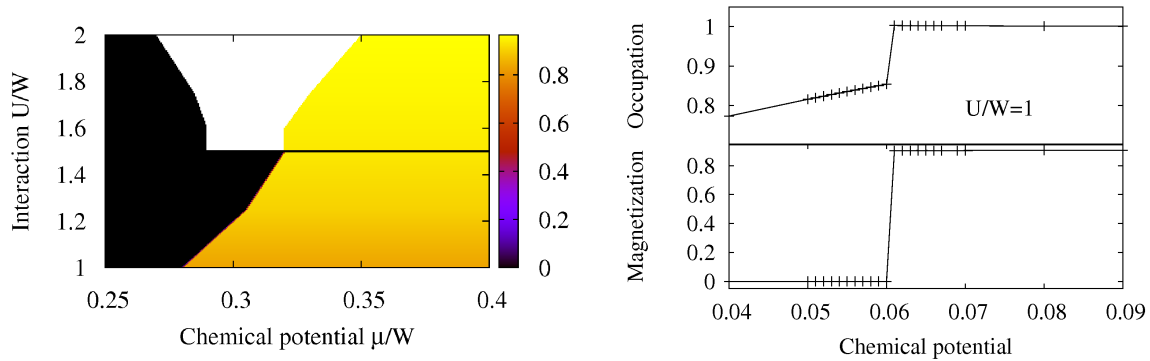}
\end{center}
\caption{Magnetic phase diagram for $t_2=0$ 
at $T/W=2\cdot10^{-4}$. Left picture: colour coded antiferromagnetic polarisation
  around half filling. The yellow part encodes the N\'eel state, black
  colour the paramagnetic state, white part the incommensurate
  phase. The black line denotes the interaction
    strength, at which in the paramagnetic phase the metal insulator
    transition would occur. The whole plot was created
by fitting of approximately 200 data points distributed in the
diagram. The right picture shows the dependence of the staggered
  magnetisation and occupation of the chemical potential $\mu$  for $U/W=1$. }\label{unfrust}
\end{figure}

Before discussing the magnetic phases within DMFT of the
system with finite $t_2$,
let us briefly review the results for the case $t_2=0$.
Figure \ref{unfrust} shows the N\'eel- and the paramagnetic 
state around half filling. 
The N\'eel state does only exist exactly at half filling. For
interaction strengths below the critical value of the paramagnetic
metal insulator transition $U_{MIT}$ (black line in the left panel) we find
phase separation between the 
N\'eel state and the paramagnetic state, which can be seen in the right
panel of figure~\ref{unfrust}. There tuning the chemical potential
leads to a jump in magnetisation and occupation. For larger values of 
the interaction $U>U_{MIT}$ there is a parameter region, where our
calculations do not converge (c.f.\ also \cite{zitzler2002}). If one 
looks at the occupation and magnetisation as function of DMFT-iteration, they show an
oscillatory behaviour with periods longer than two. Motivated by similar
previous observations \cite{peters2009} we interpret such a behaviour
as indication that 
an incommensurate spin spiral is the proper magnetic state. Note that
within a simple $AB$ lattice calculation such a spin-spiral cannot be
stabilised, and consequently
calculations do not converge in this parameter region. As we cannot determine the
nature of the magnetic order, we left this region blank in figure~\ref{unfrust}.
Apparently,
where for the paramagnet at half filling
the metal insulator transition occurs, the magnetic state
of the doped system also changes from phase separated to an incommensurate 
structure.

A ferromagnetic state, on the other hand, cannot be stabilised for the Bethe
lattice at $t_2=0$. Note that this is strikingly different from the hypercubic
lattice, where for large $U$ and small doping a Nagaoka ferromagnet occurs
\cite{zitzler2002}. The explanation is that Nagaoka's state needs closed loops
on the lattice, which are available for the hypercube (leading to
the exponential tails),
but are absent for the Bethe lattice. Thus, although in DMFT only the DOS enters
the calculations, subtle differences in the structure and support may matter
quite strongly for certain aspects of the physics.

As the DOS is particle-hole symmetric for $t_2=0$,
the phase diagram is completely symmetric with respect to half
filling. 
\subsection{$0<t_2\le 1/4 t_1$}
As $t_2$ becomes finite the DOS becomes asymmetric and consequently 
the magnetic phase diagram becomes asymmetric with respect to half filling, too.
However, for sufficiently small values of $t_2$ it will still look
very similar to figure \ref{unfrust}, 
with two notable exceptions: For the hole doped side of the phase diagram,
the incommensurate magnetic phase sets in at smaller values of the interaction,
while on the electron doped side it starts for larger values of the
interaction. Thus, for electron doping, phase separation between the antiferromagnetic
state at half filling and the paramagnetic state at $n>1$ prevails for stronger
interaction strengths. Already for $t_2/t_1=0.2$ we found no incommensurate
phase on the electron doped side for $U/W<3$. As already stated
previously 
\cite{duffy1997,hofstetter1998,chitra1999,zitzler2004,peters2009}, 
in order to stabilise the antiferromagnetic phase for a finite
next-nearest neighbour hopping one needs 
a finite interaction strength $U_c>0$.
\subsection{$1/4 t_1<t_2\le t_1$}
\begin{figure}[htp]
\begin{center}
\includegraphics[width=\textwidth,clip]{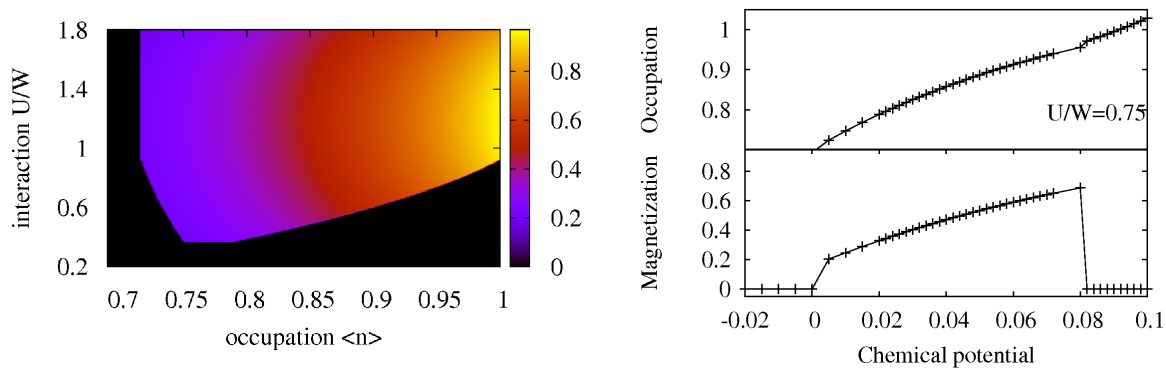}
\end{center}
\caption{Magnetic phase diagram for
  $t_2/t_1=0.8$ and $T/W=2\cdot10^{-4}$. The left plot shows the
  staggered magnetisation versus occupation and interaction
  strength. Notice that the antiferromagnetic phase sets in first away from half
  filling for increasing interaction. The right panel shows
  occupation and magnetisation for one interaction strength for which
  the half filled solution is a paramagnetic metal.}\label{anti04}
\end{figure}

For $1/4 t_1<t_2\le t_1$ one obtains according to Figure\ \ref{DOS}
a strongly asymmetric DOS showing a square-root singularity at the
lower band edge.  
Here we expect, and observe, a radically different phase diagram.
As can be seen for $t_2/t_1=0.8$ in figure \ref{anti04}
the N\'eel state can now be hole doped and does extend to large values of
the doping, i.e.\ strong frustration seems to stabilise
the N\'eel state. The incommensurate phase, on the other hand, completely
vanished from the phase diagram.
If one inspects figure \ref{anti04} more closely, one sees that the
antiferromagnetic state actually sets in away from half filling for increasing
interaction strength. At half filling we find for this values of
interaction a paramagnetic metal. On the electron doped side, we only
find a paramagnetic state, which is still phase separated to the
antiferromagnetic state at half filling.

As discussed in our previous work for half filling \cite{peters2009},
for very large $t_2/t_1>0.96$ there appears a new 
phase
which, motivated by a $120^\circ$ order expected for a classical spin system
at this level of frustration, we interpreted as such a $120^\circ$ order.
\begin{figure}[htp]
\begin{center}
\includegraphics[scale=0.8,clip]{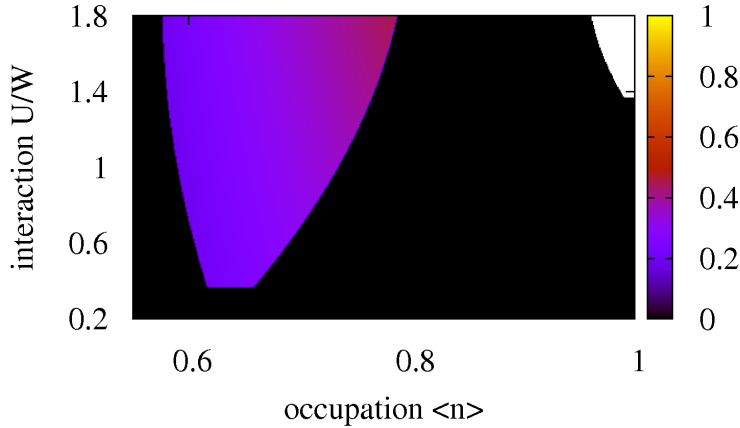}
\end{center}
\caption{Magnetic phase diagram for
  $t_2/t_1=1$ and $T/W=2\cdot10^{-4}$. There is still an
  antiferromagnetic state, which is only stable away from half filling. The
white area represent again a region of non convergent DMFT calculations
(see also text and \cite{peters2009}).}\label{anti05}
\end{figure}
Figure \ref{anti05} shows the phase diagram for
$t_2=t_1$, i.e.\ a with respect to antiferromagnetic order fully frustrated
spin system. The parameter region for large interaction left blank denotes
precisely this $120^\circ$ state, which also can be hole doped. What is
most remarkable and rather mysterious, even for the 
fully frustrated system we found a stable N\'eel state for 
fillings between $0.55<n<0.8$. 
To ensure that this result is not a numerical artifact, we performed
several calculations at different temperatures and with different
NRG parameters like discretization or states kept.
However, for low enough temperatures we always found this antiferromagnetic island.
We will come back to this point in the last section.
\section{Ferromagnetism}
As already mentioned in the introduction, while antiferromagnetism is
the ``natural'' order occurring in single-band systems as studied here,
ferromagnetic order is usually only obtained under more restrictive conditions.
In this section we therefore want to focus on possible ferromagnetic solutions
in our system. 

One of the first heuristic treatments
of metallic ferromagnetism was by E.\ Stoner \cite{stoner1938}. He gave
the criterion $UD_F>1$ for stabilising ferromagnetism, 
where $U$ is the value of the on site Coulomb interaction and $D_F$ is
the value of the density of states at the Fermi level. 
Already in this criterion one sees that ferromagnetism is created by
the interplay of the kinetic energy, characterised by $D_F$, and the Coulomb
interaction, characterised by $U$.
A rigorous result was obtained by Nagaoka \cite{nagaoka1966}, who
proved the existence of 
ferromagnetism at $U=\infty$ and ``one hole'' for certain
lattices. 
 
In the beginning of the 1990's, Mielke and Tasaki proved the existence of
ferromagnetism under certain conditions on the dispersion, known as ``flat band ferromagnetism''
\cite{mielke1991,tasaki1992}.  
Here the
ferromagnetic groundstate appears due to a dispersionless (flat) lowest
lying band. This flat band introduces a huge
degeneracy of the groundstate at $U=0$, which is lifted by
the Coulomb interaction.
A nice overview about this topic and other rigorous results for
ferromagnetism can be found in \cite{tasaki1998}.

\begin{figure}[htp]
\begin{center}
\includegraphics[width=0.8\textwidth,clip]{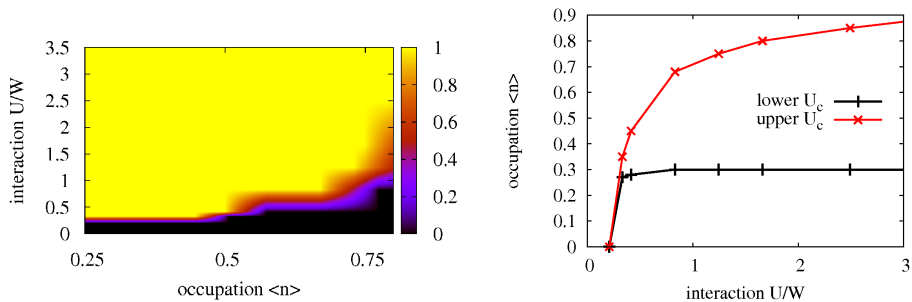}
\caption{(left panel) Ferromagnetic polarisation for $T/W=2\cdot 10^{-4}$ and
  $t_2/t_1=0.6$ for different occupations and interaction
  strengths The colour map plot was created by fitting numerical
  data. (right panel) The upper and the lower occupation for 
  stabilising ferromagnetism at different interaction
  strengths. The symbols show the interaction strengths, where
  numerical simulations were done.}\label{ferro003} 
\end{center}
\end{figure}

Remembering the singularity in the DOS for $t_2/t_1>0.25$
(see figure\ \ref{DOS}), the situation present in our system is very
similar to the ``flat band'' scenario.
Former studies for an asymmetric DOS 
\cite{wahle1998,ulmke1998,arita2000,pandey2007}
already showed the existence of ferromagnetism in such a situation. 
Consequently, we have to expect
ferromagnetism in our system, too. 
Indeed, Figure\ \ref{ferro003} shows the ferromagnetic
polarisation $\frac{n_\uparrow-n_\downarrow}{n_\uparrow+n_\downarrow}$
colour encoded over the occupation $n_\uparrow+n_\downarrow$ and the
interaction strength at low temperature ($T/W=2\cdot 10^{-4}$). The NNN
hopping for this system is $t_2/t_1=0.6$.
One sees that the singularity in the DOS alone cannot create
ferromagnetism. Here one again needs a finite interaction strength of
approximately $U/W\approx 0.3$, which however is a realistic number
for transition metal compounds of both the 3d and 4d series. 
In the right panel of figure\
\ref{ferro003} we depict the lower and upper critical occupation 
between which the ferromagnetic state is stable as function of the
interaction strength.
Below the lower critical occupation, our DMFT simulations do not
converge independent of the interaction strength. We believe that
this is a numerical problem due to the singularity in the DOS: If the
Fermi level lies very close to the singularity, the slope of the DOS at
the Fermi level is very large. Small differences in the position of
structures in the interacting Green's function will consequently have a great
influence. We however cannot rule out the possibility of the existence of
another phase in this regime. The occupation number jumps in this
region between almost zero and a larger value, and cannot be stabilised. The
behaviour can only be seen at low 
temperatures and for $t_2/t_1>0.25$, where the singularity in the DOS is
sufficiently strong and not smeared by temperature broadening.

At the upper critical occupation and low interaction strengths the
system jumps from a fully polarised ferromagnet to a paramagnetic phase.
For
strong interaction the upper occupation is large enough such that
the system directly changes from a ferromagnetic state into the 
incommensurate phase or the N\'eel phase.

\begin{figure}[htp]
\begin{center}
\includegraphics[width=0.8\textwidth,clip]{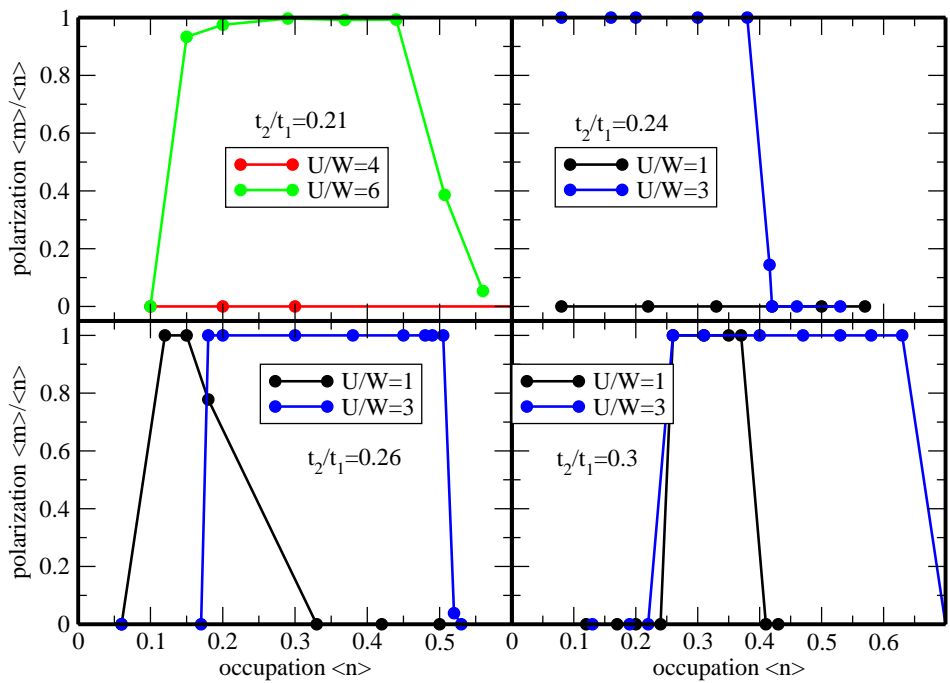}
\caption{Ferromagnetic polarisation for $T/W=2\cdot 10^{-4}$ for $t_2/t_1$
  as the singularity moves into the band. The upper panels show plots
  for DOS without singularity. Note that with increasing $t_2/t_1$
  the interaction one needs to stabilise the ferromagnet decreases.
}\label{ferrolowt2}
\end{center}
\end{figure}

As we already noted, the ``flat band'' scenario indicates that the ferromagnetic
state is intimately connected to the appearance of the van-Hove singularity
at the lower band edge.
Let us therefore look somewhat closer on the relation of the formation of a
ferromagnetic state and the appearance of the singularity in the DOS.
Figure \ref{ferrolowt2} shows the polarisation versus the occupation for
different NNN hopping $t_2/t_1$ and interaction strengths. 
The upper panels represent a situation where there is no singularity
present in the DOS. The interaction needed to stabilise a
ferromagnetic state in these systems without
singularity is strongly increased. For the case of $t_2/t_1=0.2$ we found no ferromagnetic
phase for interactions as strong as $U/W\approx 10$. As soon as
$t_2/t_1>0.25$, the critical interaction strength lies below $U/W=1$.
Increasing NNN hopping $t_2$ as well as increasing the interaction
strength favours the ferromagnetic state as the region in occupation
gets more and more extended. In the DMFT/QMC study of Wahle et
al.\ \cite{wahle1998} a 
peak at the lower band edge was enough to stabilise a ferromagnetic
phase at moderate interaction strengths. In our
calculations the tendency towards ferromagnetism dramatically
decreases for a DOS without singularity.
\section{Competition between ferromagnetism and antiferromagnetism}
A careful look at the phase diagrams reveals that 
there are parameter regions where one seemingly can obtain both
an antiferromagnetic as well as a ferromagnetic solution to the DMFT equations.
This is rather unusual because conventionally DMFT will show oscillating
behaviour if one performs a ferromagnetic calculation in a regime with
antiferromagnetic ground state and vice versa. 

To decide which of the two solutions is the thermodynamically stable one, one
has to compare their respective free energies.
As the calculations were done practically at $T=0$,
we calculate the energy of the system, given by
\begin{displaymath}
\frac{\langle H\rangle}{N}=\frac{\langle H_T\rangle}{N} +
\frac{U}{N}\sum_i\langle n_{i\uparrow}n_{i\downarrow}\rangle
\end{displaymath}
where $H_T$ is the kinetic energy and $N$ the number of sites. The interaction
term is purely local and thus can be taken from the converged impurity
calculation.

The kinetic energy on the other hand can be calculated from the expression
\begin{displaymath}
\langle H_T\rangle=\int\limits_{-\infty}^\infty d\theta\epsilon(\theta)\rho(\theta)\int\limits_{-\infty}^0d\omega
\left(-\frac{1}{\pi}\right)\Im m\frac{1}{\omega+\mu-\epsilon(\theta)-\Sigma(\omega+i\eta,\theta)}
\end{displaymath}
where $\Sigma(z,\theta)$ is the lattice self-energy, $\theta$ a suitable
variable to label the single-particle energies on the lattice under
consideration and $\mu$ the
chemical potential. Within DMFT, the lattice self-energy is
approximated by a local self-energy, i.e.\ we may set 
$\Sigma(z,\theta)=\Sigma(z)$. Furthermore, for the Bethe lattice with 
infinite coordination
$\epsilon(\theta)=t_1\theta+t_2(\theta^2-1)$ and
$\rho(\theta)=\frac{1}{2\pi}\sqrt{4-\theta^2}$ holds. Substituting $\epsilon(\theta)$
by $\epsilon$ in the integral, the resulting DOS takes on the form given in section \ref{sec:DOS}.

\begin{figure}[htp]
\begin{center}
\includegraphics[width=0.6\textwidth,clip]{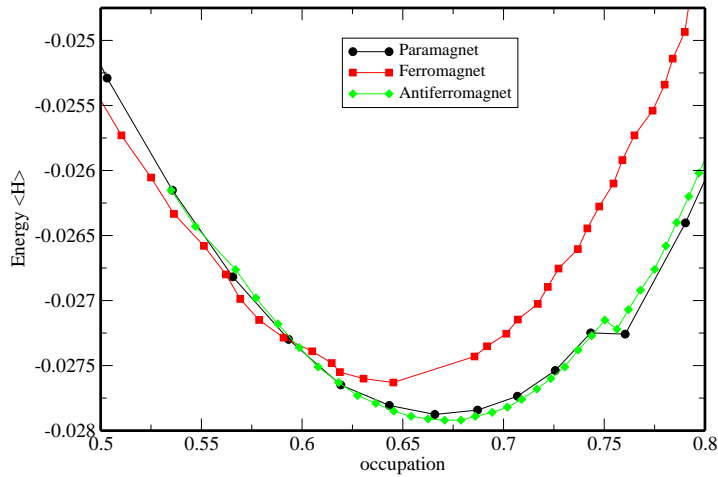}
\caption{Energies for the converged paramagnetic, ferromagnetic and
  antiferromagnetic solution for $t_1=t_2$ and $U/W=2.5$. The lines
  are meant as guide to the eye.}\label{energy}
\end{center}
\end{figure}

Since the N\'eel state is defined on an $AB$ lattice, one has to distinguish
between the inter- and intra-sublattice hopping terms, and 
the formula for the kinetic energy takes on the form
\begin{displaymath}
\langle H_T\rangle
=-\frac{1}{\pi}\Im m
\int\limits_{-\infty}^\infty
d\theta\rho(\theta)\int\limits_{-\infty}^0d\omega\!\begin{array}[t]{l}\displaystyle
\Bigl(t_1\theta
\left(G_{AB}(\omega+i\eta)+G_{BA}(\omega+i\eta)\right)+\\[5mm]
\displaystyle t_2(\theta^2-1)\left(G_{AA}(\omega+i\eta)+G_{BB}(\omega+i\eta)\right)\Bigr)\end{array}
\end{displaymath}
Note that with the definition of the matrix Green function this formula can be
put into the compact matrix form
\begin{eqnarray*}
\langle H_T\rangle &=&
-\frac{1}{\pi}\Im m\int\limits_{-\infty}^\infty
d\theta\epsilon(\theta)\rho(\theta)\int\limits_{-\infty}^0d\omega\sum_{ij}
\left[\underline{\underline{G}}(\omega+i\eta)\right]_{ij}\\
\left[\underline{\underline{G}}(\omega+i\eta)\right]_{ij} &:=&
\left(\begin{array}{cc}\zeta_\uparrow-t_2(\theta^2-1)&-t_1\theta\\-t_1\theta&\zeta_\downarrow-t_2(\theta^2-1)\end{array}\right)^{-1}_{ij}\\
\zeta_\sigma(\omega) &:=&\omega+\mu-\Sigma_\sigma(\omega+i\eta)
\end{eqnarray*}
The energies of the converged solutions for $t_2=t_1$
and $U/W=2.5$ can be seen in figure~\ref{energy}. 
The antiferromagnetic solution could be stabilised in this parameter
region for occupations
$0.55<n<0.8$. From figure
\ref{energy} it becomes clear now that the ferromagnetic state has the
lowest energy for $n<0.6$. For $0.6<n<0.75$ the
antiferromagnetic state takes over as the groundstate, but is nearly
degenerate with the paramagnetic state. 
For fillings larger than $0.8$ no staggered magnetisation can be stabilised
any more.\par
Thus, the energy calculations reveal two things. Firstly, an
antiferromagnetic N\'eel state indeed seems to form away from half filling
in the fully frustrated system. Secondly, the energy differences are extremely
small, in particular the antiferromagnet and paramagnet are de facto degenerate
over the full parameter regime where the former exists.

To understand this at first rather
irritating observation let us recall the well-known fact that
in strongly frustrated systems it is a common feature to have a large number of
degenerate groundstate configurations, which also can include 
magnetically ordered ones
\cite{tasaki1998}. 
Thus, the degeneracy of the antiferromagnet and the paramagnet
hints towards
the possibility that there may exist a larger number of other magnetically ordered states in this
parameter region.
Unfortunately
we are not able  
to search for and in particular stabilise those magnetic phases with the technique
at hand. Further
investigations using different methods to solve the DMFT equations are
definitely necessary.

\section{Conclusions}
In conclusion, we have calculated the magnetic phase diagram for the
Bethe lattice with NN- and NNN-hopping in the limit of infinite
coordination. For this purpose we have used the proper expression for the DOS
of this lattice as deduced by Kollar~et~al.~. By varying the NNN hopping
one can tune the DOS from a symmetric semi-ellipse to a very asymmetric
shape with a square-root van-Hove singularity at the lower band edge. While the electron
doped side of the phase diagram tends to phase separate between the
N\'eel state and a paramagnetic metal just like at the
particle-hole symmetric point, 
the hole doped side reveals a surprisingly rich phase diagram.

We first note that the regimes with phase separation respectively
incommensurate spin-spiral states are  replaced by a doped N\'eel state.
As expected, we need a finite interaction $U_c$ to allow the existence of
the N\'eel state, which for larger $t_2$ has its minimum at finite doping, 
i.e.\ the N\'eel state is first formed away from half filling.

In addition, with increasing NNN hopping $t_2$ a ferromagnetic phase at low fillings
can be found. For large $t_2$ and strong interaction $U$ this
ferromagnetic phase can extend to occupations $n>0.7$. The dependence
of the appearance of this phase on the parameter $t_2$ shows that it
is related to Mielke's and Tasaki's notion of ``flat-band'' ferromagnetism
rather than Nagaoka's ferromagnetism found at low \emph{doping} and $U\to\infty$
in the hypercubic lattice.

Quite amazingly, we found that for $t_2\approx t_1$ and large
enough interaction $U$
a doped antiferromagnet can also be stabilised in the same
filling region. 
Calculating the groundstate energies of both magnetic states 
and the paramagnetic solution, we find that the
ferromagnet is the ground state below 
some critical filling  $n_c$. For $n>n_c$, the N\'eel state and the paramagnet
are degenerate within numerical accuracy and lower than the ferromagnet.

Finding both magnetic states stable within DMFT and the near
degeneracy of them could be an effect of the strong
frustration,  
where a large number of degenerate or nearly degenerate groundstates is a common
feature. This would also mean that in this parameter region in our
model more magnetically ordered states should be observable.
As we are however only able to
look for homogeneous or N\'eel states, this is only speculative, nevertheless
motivating further studies of magnetic order in the single-band Hubbard
model with different methods to solve the DMFT equations. However, for these
studies the Bethe lattice may not be a suitable choice any more, as the definition of 
a wave vector $\vec{Q}$ to identify the various possible spin structures is
not possible here. 
\begin{ack}
We want to thank A.\ Honecker for helpful discussions. One of us (TP)
acknowledges the hospitality of the Racah Institute of Physics at the Hebrew University
of Jerusalem. This work was
supported by the DFG through  
PR298/10. Computer support was provided by the 
Gesellschaft f\"ur 
wissenschaftliche Datenverarbeitung in G\"ottingen and the
Norddeutsche Verbund f\"ur Hoch- und H\"ochstleistungsrechnen.
\end{ack}

\end{document}